# Wigner Distribution Function and Entanglement of Quantum Optical Elliptical Vortex

Abir Bandyopadhyay[a, b], Shashi Prabhakar[b] and R. P. Singh[b]

[a]Hooghly Engineering and Technology College, Hooghly 712103, India. abir@hetc.ac.in
[b]Quantum Optics & Quantum Information Group, Theoretical Physics Division,
Physical Research Laboratory, Navrangpura, Ahmedabad – 380009, India

**Abstract.** We calculate the Wigner (quasi)probability distribution function of the quantum optical elliptical vortex (QEV) generated by coupling squeezed vacuum states of two modes. The coupling between the two modes is performed by using beam splitter (BS) or a dual channel directional coupler (DCDC). The quantum interference due to coupling between the two modes promises the generation of controlled entanglement for quantum computation and quantum tomography. We compute the entanglement of such QEV formed by non-classical radiation field, using Wigner function. We report a critical squeezing parameter above which the entanglement is less for higher vorticity, which is counter intuitive.

## I. INTRODUCTION

Phase space, a fundamental concept in classical mechanics, remains useful when passing to quantum mechanics. To study the properties of quantum states a number of (quasi)probability distributions have been introduced in quantum mechanics, on the line of probability density distribution functions in classical systems [1]. They provide a complete description of quantum systems at the level of density operators, though not at the level of state vectors. However, among all the (quasi)probability distributions, the Wigner function stands out, as it is real, nonsingular, yields correct quantum-mechanical operator averages in terms of phase-space integrals, and possesses positive definite marginal distributions [2]. The Wigner distribution function has come to play an ever increasing role in the description of both coherent and partially coherent beams and their passage through first order optical systems [3]. Once the Wigner distribution is known, the other properties of the system can be calculated from it.

Entanglement is fundamental in determining the usefulness of a quantum state in quantum information protocols. Entangled states play a central role in quantum key distribution, superdense coding, quantum teleportation, and quantum error correction [4], which cannot be realized classically. While a major part of the effort in quantum information theory has been in the context of systems with finite number of Hilbert space dimensions, more specifically the qubits, recently there has been much interest in the canonical continuous cases [5]. Gaussian states, which are part of continuous variable case, are much better studied and methods have been devised to measure the entanglement for such states [6, 7]. The measure of entanglement for a Gaussian state is best characterized by the logarithmic negativity [8], evaluated in terms of the symplectic eigenvalues of the covariance matrix [7], and computed from the Wigner function.

Circular optical vortex beams with helical wave front can be produced in a controlled manner using methods such as computer-generated hologram (CGH), cylindrical lens mode converter, and spiral phase plate [9]. Optical vortices have drawn a great attention in the last two decades and these have prompted to start a new branch in physical optics known as singular optics [10]. For a vortex, the OAM follows the same numbers as vorticity or topological charge of the vortex i.e. an optical vortex of topological charge $m$, carries an OAM of $m\hbar$ per photon. Because of their specific spatial structure and associated orbital angular momentum (OAM), they find applications in the field of optical manipulation, optical communication, quantum information and computation [11]. However, most of the work relating to optical vortex deals with classical vortex that involves classical electromagnetic field. It is rare to find literature that takes into account the vortex formed by quantized radiation field. Agarwal *et al.* have generated quantum vortex from two separate squeezed vacuum modes (TSSM) [12]. They have shown that a two mode state under the linear transformation belonging to the SU(2) group may lead to a vortex state under special conditions [13]. However, they considered the special case of 50/50 superposition of two orthogonal fields, producing the vortex as a Fock state in circular basis [12].

Recently we have generated quantum optical elliptic vortex (QEV) by coupling squeezed vacuum of two modes (TSSM) with the help of a beam splitter (BS) or a dual channel directional coupler (DCDC) [14]. Both the components, BS and DCDC, find practical applications in optical coherence tomography [15]. In this work, we compute the Wigner function and with help of that discuss how the coupling of modes could be used to generate controlled entanglement for application to quantum computation and quantum information. We show this by quantifying the entanglement in terms of the logarithmic negativity. The paper is organized as follows. In section II we describe the QEV and study the properties of the Wigner function of the QEV. In section III we discuss the definition of entanglement for Gaussian input states and compute it with the help of Wigner function, computed in Sec. II. We find the entanglement for QEV for a certain choice of parameters and discuss the result. We find a critical value of the squeezing parameter, above which higher vorticity produces less entanglement. We conclude our results in section IV.

## II. WIGNER DISTRIBUTION OF QUANTUM OPTICAL ELLIPTICAL VORTEX

### (A) Generation of Quantum Optical Elliptical Vortex (QEV), from two Squeezed Vacuum, with Beam Splitter (BS) or Dual Channel Directional Coupler (DCDC)

It is possible to couple two beams by using a beam splitter (BS) or a dual channel directional coupler (DCDC) [12, 13, 16, 17]. The BS is used by many authors as an entangler [16]. For two-mode states characterized by the annihilation operators $a_1$ and $a_2$, a coupling transformation can be generated by evolution under a Hamiltonian of the form $H = g(a_x^\dagger a_y \, e^{i\varphi} + \text{h.c})$. While Kim et al. examined the question of the generation of entangled states by a beam splitter using Fock states as input fields [17], Agarwal and Banerjee constructed circular vortex state using the above mentioned Hamiltonian, described for BS/DCDC, and studied the properties of its entropy [13]. Considering output operators $a_i^\dagger(out)$ are generated by the unitary transform $\mathcal{U}^\dagger a_i^\dagger(in)\mathcal{U}$, for the above Hamiltonian

$$\begin{bmatrix} a_x^\dagger(out) \\ a_y^\dagger(out) \end{bmatrix} = \begin{bmatrix} A_x & A_y \\ A_y^* & A_x^* \end{bmatrix} \begin{bmatrix} a_x^\dagger(in) \\ a_y^\dagger(in) \end{bmatrix}, \qquad (1)$$

where $\mathcal{U} = e^{-iH}$. $A_i$ denote transmitivity and reflectivity of the BS respectively and satisfy the relations $|A_x|^2 + |A_y|^2 = 1$, and $A_x^* A_y + A_y^* A_x = 0$. Mixing of equal amount ($A_x = A_y$) generates a circular vortex which has been dealt quantum mechanically elsewhere [12, 13]. In the cases where the coefficients of mixing ($A_i$) are not equal, they produce an elliptical vortex. As the asymmetry becomes larger and larger, more "which path" information is available, and the quantum interference effect is correspondingly diminished. Somewhat surprisingly, this reduced interference has been found to be extremely useful in a number of quantum information processing applications in linear optics, such as, quantum computing gates [18] and quantum cloning machines [19].

We consider two separate squeezed vacuum modes (TSSM) as our input states and couple them through a BS or DCDC. If we look at any of the output states after *m* times the operation is performed, it is given by [14],

$$|\Psi_{qev}\rangle = \mathcal{N} \left[ \eta_x a_x^\dagger \pm i\, \eta_y a_y^\dagger \right]^m S_x(\zeta_x)\, S_y(\zeta_y) |0,0\rangle, \qquad (2)$$

where $\mathcal{N}$ is the normalization constant and $S_i(\zeta_i) = exp(\zeta_i^* a_i^{\dagger 2} - \zeta_i a_i^2)$ are the usual squeezing and displacement operators corresponding to *x* and *y* modes/directions. We call these states quantum optical elliptical vortex (QEV). The term in square bracket, generated by BS/DCDC, is responsible for the elliptical vortex. If we put $\eta_x = \eta_y = 1$, and $\zeta_x = \zeta_y = \zeta$ (real), it reduces to the quantum optical circular vortex state (QCV): $|\Psi_{qcv}\rangle$. QCV has been studied in detail in [12] using Husimi Q function. When $\eta_x \neq \eta_y$, it refers to elliptical vortex. The parameters in the generator of the vortex term, $\eta_i$, are trivially connected to the reflection and transmission of the BS, or to the coupling ratios for DCDC, as described in Eq. (1).

### (B) Computation of Wigner (Quasi) Probability Distribution Function of the QEV

Following the mathematical treatments of [12], with the choice of the parameters, $\eta_i = 1/(\sqrt{2}\,\sigma_i)$, for $i = (x, y)$, one can calculate the spatial distribution of QEV state as [14]

$$\Psi_{qev}(x, y) = \sqrt{\frac{2^{(m-2)}}{\sigma_x \sigma_y \Gamma(m+\frac{1}{2})\sqrt{\pi}}} \left[ \frac{x}{\sqrt{2}\sigma_x} \pm i \frac{y}{\sqrt{2}\sigma_y} \right]^m exp\left[ -\frac{1}{2} \left\{ \left(\frac{x}{\sigma_x}\right)^2 + \left(\frac{y}{\sigma_y}\right)^2 \right\} \right], \qquad (3)$$

where, $\sigma_i = exp(2\zeta_i)$. The distribution $|\Psi_{qev}(x,y)|^2$ is shown to have elliptic vortex structure [14]. Inverting the ratio $\sigma_x/\sigma_y$ rotates the ellipse by $\pi/2$.

Now, we change our variables to scaled ones: $X_1 = x/\sigma_x$, $Y_1 = y/\sigma_y$, $P_{x_1} = (\sigma_x p_x)/\sqrt{2}$, $P_{y_1} = (\sigma_y p_y)/\sqrt{2}$, $X_2 = (\sigma_y x)/(2\sigma_x)$, $Y_2 = (\sigma_x y)/(2\sigma_y)$, $P_{x_2} = (\sigma_y^3 p_x)/\sqrt{2}$, $P_{y_2} = (\sigma_x^3 p_y)/\sqrt{2}$. Following the treatment in [20], it allows us to compute the four dimensional Wigner function for the state $|\Psi_{qev}\rangle$ in a compact fashion as,

$$W(x, y, p_x, p_y) = K \exp\left[-\left(X_1^2 + Y_1^2 + P_{x_1}^2 + P_{y_1}^2\right)\right] L_m^{-1/2}\left[\frac{\left(P_{x_2} + P_{y_2} - X_2 - Y_2\right)^2}{\sigma_x^2 + \sigma_y^2}\right] \quad (4)$$

where, $L_m^{-1/2}[\;\;]$ is associated Laguerre polynomial (ALP), and, $K = \frac{2^{(m-4)} m!}{\pi\sqrt{\pi}\Gamma(m+\frac{1}{2})}\left[-2(\sigma_x^2 + \sigma_y^2)\right]^m$. Note that in the Eq. (4), the changed variables in the Gaussian term are different from the changed variables in the argument of the ALP term. In case of a circular vortex the hole and vortex terms factor out as a product $r^{2m} L_m^0$, along with the Gaussian term. In the present case, the usual Gaussian term is factored out nicely, but the hole term ($r^{2m}$) is not separated out from the Laguerre term. We notice that it is embedded in the ALP term, which ensures that the elliptical vortex may be expressed as a combination of circular vortices from 0 to $m$.

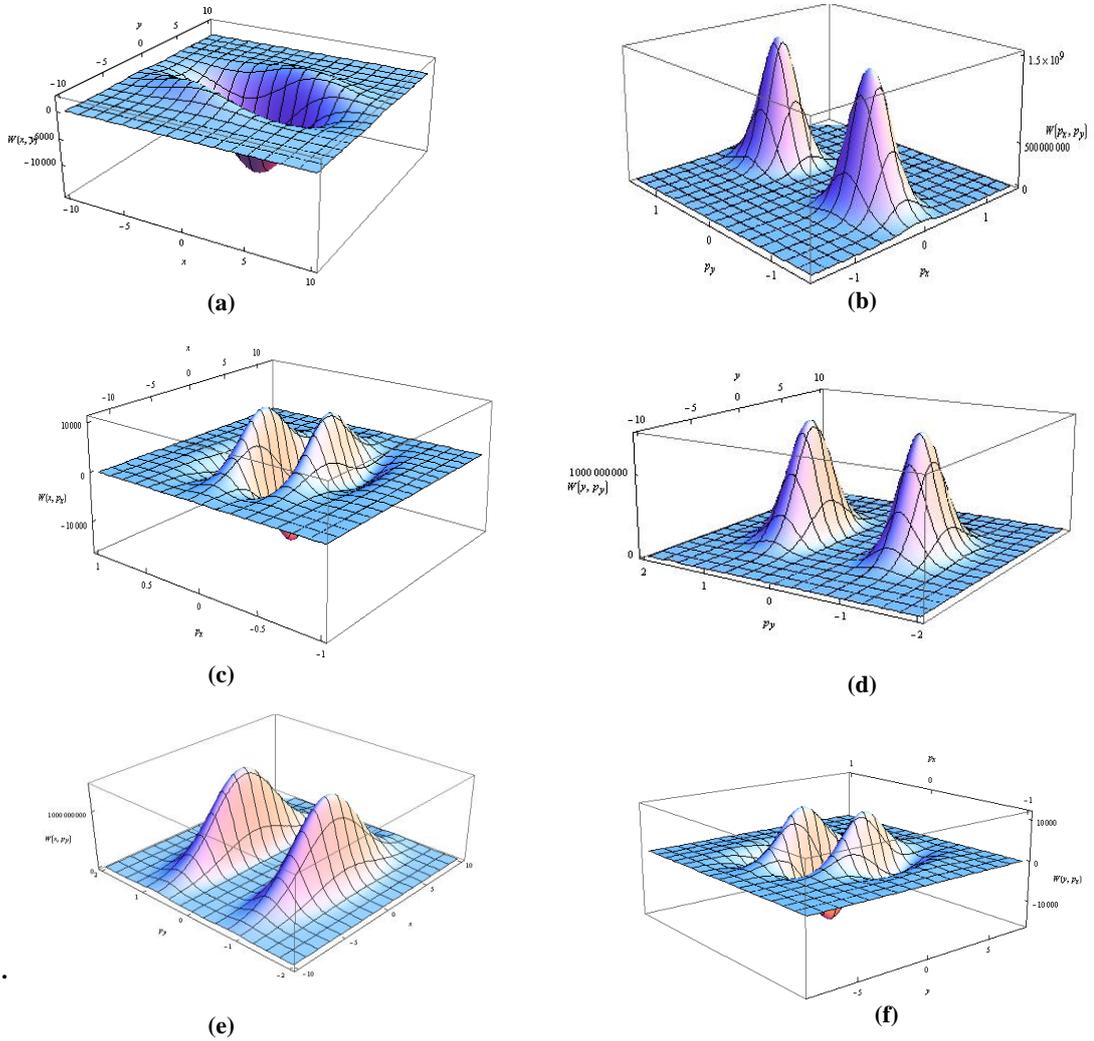

**FIGURE 1.** Plot of Wigner function (in arbitrary units) for $m = 3$, with $\sigma_x = 5$, $\sigma_y = 3$ as a function of (a) $x$ and $y$ (b) $p_x$ and $p_y$ (c) $x$ and $p_x$ (d) $y$ and $p_y$ (e) $x$ and $p_y$ (f) $p_x$ and $y$  [14].

Next we study the four dimensional Wigner function, Eq.(4), in detail. It can be reduced to two variables by choosing the other two to be zero. For such choices the Wigner function can be written as six reduced functions, each involving only two variables (2D),

$$W(x,y)_{p_y=0}^{p_x=0} = K\, exp[-(X_1{}^2 + Y_1{}^2)]L_m^{-1/2}\left[\frac{(Y_2+X_2)^2}{\sigma_x^2+\sigma_y^2}\right], \tag{5a}$$

$$W(p_x,p_y)_{y=0}^{x=0} = K\, exp\left[-\left(P_{x_1}{}^2 + P_{y_1}{}^2\right)\right]L_m^{-1/2}\left[\frac{(P_{y_2}+P_{x_2})^2}{\sigma_x^2+\sigma_y^2}\right], \tag{5b}$$

$$W(x,p_x)_{p_y=0}^{y=0} = K\, exp[-(X_1{}^2 + P_{x_1}{}^2)]L_m^{-1/2}\left[\frac{(P_{x_2}-X_2)^2}{\sigma_x^2+\sigma_y^2}\right], \tag{5c}$$

$$W(y,p_y)_{p_x=0}^{x=0} = K\, exp\left[-\left(Y_1{}^2 + P_{y_1}{}^2\right)\right]L_m^{-1/2}\left[\frac{(P_{y_2}-Y_2)^2}{\sigma_x^2+\sigma_y^2}\right], \tag{5d}$$

$$W(x,p_y)_{y=0}^{p_x=0} = K\, exp\left[-\left(X_1{}^2 + P_{y_1}{}^2\right)\right]L_m^{-1/2}\left[\frac{(P_{y_2}-X_2)^2}{\sigma_x^2+\sigma_y^2}\right], \tag{5e}$$

$$W(y,p_x)_{p_y=0}^{x=0} = K\, exp[-(Y_1{}^2 + P_{x_1}{}^2)]L_m^{-1/2}\left[\frac{(P_{x_2}-Y_2)^2}{\sigma_x^2+\sigma_y^2}\right], \tag{5f}$$

Though all the six relations look alike up to the Gaussian terms, a careful look provides subtle differences between them. While the numerators of the arguments of the ALP terms of the first two are square of the *sum* of the two variables, in the other four relations the same terms are square of the *difference* of the variables. Due to the presence of a square in all the six terms, cross terms between the two variables appear in the Wigner functions. The first two relations, Eqs. (5a,b), provide information about the cross correlation between the same quadratures of two different modes. The relations in Eqs. (5c,d) provide the information about the cross correlation between the two quadratures of the same mode. The last two relations, Eqs. (5e,f), show quantum interference in the cross correlation of different quadratures of two different modes.

We study the properties of the functions in Eqs. (5a-f) by plotting them in Fig (1) [14]. We choose $\sigma_x = 5$, $\sigma_y = 3$, and $m = 3$ for all the Figs. (1a-f), though we also discuss the observations for other values of the parameters. If $\sigma_i$s are interchanged, then the figures rotate by $\pi/2$. Knowing the fact that the changed variables involved in the Gaussian term are different from the changed variables in the argument of the ALP term, with the observations in the last paragraph, it is expected to observe different distribution patterns. In Fig. (1a) the ellipticity and vortex structure are observed for the Wigner function of two space coordinates. However, in the Fig. (1b), the Wigner function of the momentum of the different modes breaks up to two separate elliptic Gaussian functions. In the phase space of the *x* mode, Fig. (1c), new structures start showing up beyond the core vortex. The number of minima matches with the value of *m* with *m*+1 maxima. The outermost maxima are not very clear and need to be observed carefully for the set of parameters chosen. For even *m*, it is observed that there is an even number of minima with *m*-1 maxima in between. In Figs. (1d) and (1e) the Wigner functions in the phase space of the second mode (y, $p_y$) and the cross phase space of *x* and $p_y$ show similar squeezed Gaussian structure. In Fig. (1f), the dependence on *y* and $p_x$ shows similar behavior as in Fig. (1c). The asymmetry in the state under consideration is the reason for such similarity in the plots.

## ENTANGLEMENT FOR THE QEV

In this section we study the variation of entanglement of the QEV states with respect to the change in one of the squeezing parameters. The measure of entanglement has been done in terms of logarithmic negativity, which is well defined for the Gaussian states. We focus on our method for generating the vortex state by the propagation of light through a BS or coupled waveguides (DCDC), the unitary operations, which currently are used in quantum architectures and quantum random walks. Here, we would like to draw attention to the lemma: "*If U is a unitary map corresponding to a symplectic transformation in the phase space, i.e. if U = exp{−iH} with Hermitian H and at most bilinear in the field operators, then δA[UρU†] = δA[ρ]*" [21]. The proof of the lemma ensures that single-mode displacement and squeezing operations, as well as two-mode evolutions as those induced by a beam splitter or a parametric amplifier; *do not* change the Gaussian character of a quantum state. As the generalized vortex states, we considered, are generated only by such operations, it qualifies to be Gaussian. Note that since the first statistical moments can be arbitrarily adjusted by local unitary operations, it does not affect any property related to entanglement or mixedness and thus the behavior of the covariance matrix *Σ* is all important for the study of the entanglement. Therefore, the logarithmic negativity $E_N$ [8], a quantity evaluated in terms of the symplectic eigenvalues of the covariance matrix *Σ* [7], and measure of the entanglement for a Gaussian state, can be applied to measure the entanglement for the QEV states. The elements of the covariance matrix *Σ* are given, in terms of conjugate observables, in the symmetrized form,

$$\Sigma = \begin{bmatrix} \alpha & \mu \\ \mu^T & \beta \end{bmatrix}, \tag{6}$$

with $\alpha = \begin{bmatrix} \langle x^2 \rangle & \langle \frac{xp_x+p_xx}{2} \rangle \\ \langle \frac{xp_x+p_xx}{2} \rangle & \langle p_x^2 \rangle \end{bmatrix}$, $\beta = \begin{bmatrix} \langle y^2 \rangle & \langle \frac{yp_y+p_yy}{2} \rangle \\ \langle \frac{yp_y+p_yy}{2} \rangle & \langle p_y^2 \rangle \end{bmatrix}$, and, $\mu = \begin{bmatrix} \langle \frac{xy+yx}{2} \rangle & \langle \frac{xp_y+p_yx}{2} \rangle \\ \langle \frac{yp_x+p_xy}{2} \rangle & \langle \frac{p_xp_y+p_yp_x}{2} \rangle \end{bmatrix}$. The structure of $\Sigma$ ensures that it is the transpose of itself ($\Sigma^T = \Sigma$). The symmetric operator averages in the matrix elements of $\Sigma$ are calculated from the Wigner function using the relation $\langle \hat{O} \rangle = \iint_{-\infty}^{\infty} dx\, dp_x \iint_{-\infty}^{\infty} dy\, dp_y\, \hat{O}\, W(x,y,p_x,p_y)$.

The condition for entanglement of a Gaussian state is derived from the Peres-Horodecki positive partial transpose (PPT) criterion [7], according to which the smallest symplectic eigenvalue $\nu_<$ of the transpose of matrix $\Sigma$ should satisfy $\nu_< < 1/2$, where, $\nu_< = \min[\nu_+, \nu_-]$. The eigenvalues are given by: $\nu_\pm^2 = \left(\Delta(\Sigma) \pm \sqrt{\Delta(\Sigma)^2 - 4\text{Det}\Sigma}\right)/2$, where $\Delta(\Sigma) = \text{Det}(\alpha) + \text{Det}(\beta) - 2\text{Det}(\mu)$. Thus according to the condition, when $\nu_< \geq 1/2$ Gaussian states become separable. The corresponding quantification of entanglement is given by the logarithmic negativity $E_N$ [22] defined as,

$$E_N = \max[0, -\ln(2\nu_<)] \tag{7}$$

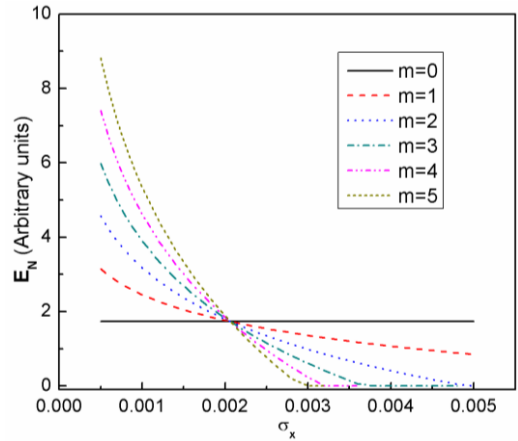

**FIGURE 2.** Plot of entanglement $E_N$ (in arbitrary units) vs. $\sigma_x$, for different topological charge (*m*) of the QEV. Note the finite constant entanglement for *m* = 0 (explained in the text) and *critical point* of squeezing parameter.

We study the dependence of entanglement on squeezing parameter. We report a *critical* squeezing parameter, above which, higher AM means lower entanglement. We have computed the entanglement for a choice of parameters $\sigma_y = \sqrt{5\sigma_x}$. In terms of $\zeta_i$, the relationship is linear: $\zeta_y = \frac{\ln 5}{4} + \frac{\zeta_x}{2}$. We have plotted the entanglement, $E_N$, in Fig. (2), for *m* = 0 to 5 i.e. for different orders of the vortex, as a function of $\sigma_x$. First of all, we analyze our observation for *m* = 0. For this state, we observe entanglement, which is counter intuitive. *m* = 0 means that no vortex is formed, thus there should be no entanglement, as it is TSSM state. We have tried to explain the reason for this apparent contradiction. It is definitely true that if the two squeezing parameters are random, then the state would be separable. However, it is possible that due to our (or, logically speaking, any) choice of a specific relation between the squeezing parameters, some sort of entanglement is imposed. Thus we argue that the constant entanglement, generated in our computation, is due to the specific choice of squeezing parameter. The observation of the constant value of the entanglement supports the logic that as it is generated with some fixed relationship, it remains constant. We have verified the fact that some other fixed relationship produces entanglement with some other constant value. However, the other dependencies of the parameters will be considered in future correspondences, if found with interesting features. The important observation in Fig. (2) is that the above a critical point, $\sigma_x = 0.002$ ($\zeta_x = -3.1073 = 3.1073\, e^{i\pi}$), the higher OAM corresponds to lower entanglement. The squeezing parameter $\zeta_i$ is complex in general. However, most of the studies consider only real positive values of the parameter. We report our work in the complex domain of the parameter from the expression of $\zeta_x$ mentioned above, where the negativity is realized in the phase factor of the complex squeezing parameter. The negative value of the squeezing parameter corresponds to the sub-Poissonian photon number distribution.

## CONCLUSIONS

To conclude, we used the two well known mechanisms (BS and DCDC) of coupling the two squeezed vacuum modes to generate a quantum optical elliptical vortex (QEV). We have computed and studied the properties of the Wigner (quasi)probability distribution function of QEV. We have argued that the QEV is a Gaussian state as squeezing or coupling between the two modes do not change this property. Thus the entanglement follows the Peres-Horodecki PPT criterion and defined in terms of the logarithmic negativity of the lowest eigenvalue of the covariance matrix. We have computed the entanglement of such QEV state with the four dimensional Wigner distribution function, which is used to find out the covariance matrix and therefore, the logarithmic negativity of its minimum eigenvalue. We show that by changing the squeezing parameter one can control the entanglement. We observed a critical point above which the increase in vorticity decreases the entanglement.

## ACKNOWLEDGMENTS

AB acknowledges the Associateship at PRL.